\documentclass[10pt,letterpaper,twocolumn]{revtex4}%
\usepackage{amsmath}
\usepackage{amsfonts}%
\usepackage{amssymb}%
\usepackage{graphicx}

\begin{document}
\title{Fast and slow light in zig-zag microring resonator chains}
\author{P. Chamorro-Posada} 
\affiliation{Dpto. de Teor\'{\i}a de la Se\~nal y Comunicaciones e Ing. Telem\'atica, Universidad de Valladolid, ETSI Telecomunicaci\'on, Campus Miguel Delibes s/n, 47011 Valladolid, Spain}
\author{F.J. Fraile-Pelaez}
\affiliation{Dpto. de Teor\'{\i}a de la Se\~nal y Comunicaciones, Universidad de Vigo, ETSI Telecomunicaci\'on, Campus Universitario, Vigo, Spain}

\begin{abstract}We analyze fast and slow light transmission in  a zig-zag microring resonator chain.  This novel device permits the operation in both regimes.  In the superluminal case, a new ubiquitous light transmission effect is found whereby the input optical pulse is reproduced in an almost simultaneous manner at the various system outputs.  When the input carrier is tuned to a different frequency, the system permits to slow down the propagating optical signal.  Between these two extreme cases, the relative delay can be tuned within a broad range.\end{abstract}

\maketitle

The ability to control the transmission delay of optical signals is a major
{technological} target pursued for the next generation of
high-performance photonic communications networks \cite{Boyd2006}. Different
physical mechanisms can provide the required control\cite{Parra2007}. The use
of microring resonator chains \cite{Heebner2002,Fraile2007,Liu2008} is
particularly appealing since the integration of large structures in a
small-size optical chip \cite{Xia2007} may be the key to achieve the high
degree of control demanded in optical transmission systems \cite{Boyd2006}.

Previous research on microring resonator chains for the control of the
{propagation} delay has focused mainly on two types of
structures\cite{Scheuer2005}: Side Coupled Integrated Spaced Sequence of
Resonators (SCISSOR) and Coupled Resonator Optical Waveguides (CROW). We
analyze a novel type of structure consisting of a Zig-Zag Ring Resonator
(ZZRR) chain. This is shown to provide a group delay which can be controlled
by {tuning} the structure {using} previously reported techniques
\cite{Blair2006,Liu2008} in {a} broad range of values. In the two
extremes, we find fast and slow light. A novel type of fast light propagation
in which the input signal is placed at several outputs in a nearly
{simultaneous} manner. Potential applications of the new effects and
device are briefly discussed.

\begin{figure}[tbh]
\centerline{\includegraphics[width=8cm]{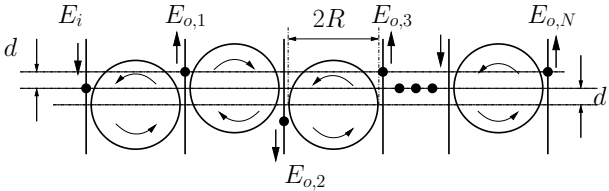}}\caption{Geometry of the zig-zag microring resonator chain. $E_i$ is the input field and $E_{o,l}$ is the $l$th output.} \label{estructura}
\end{figure}

Figure \ref{estructura} displays the topology of the physical system under study. Propagation
is assumed dispersionless for each waveguide section. The total ring length is
$L=2\pi R$ and we use for reference in the {analysis of} propagation
phenomena the length of one half-ring $L/2$ for which $\tau_{0}=L/(2v)$ is the
transmission delay, where $v$ is the (phase and group) velocity and $g$ is
the half-ring amplification ($g>1$) or attenuation ($g<1 $) factor for active
and passive structures, respectively. All the coupling constants are assumed
to have the same value of $t$ and $r=\sqrt{1-t^{2}}$. A {physical}
length $d$ is {assigned} to the coupling sections which, in turn, is
responsible for the zig-zag geometry of the structure.

The transfer function, as a function of a normalized frequency $\Omega
=\tau_{0}\omega$, is defined by the relation $E_{o,k}(\Omega)=H_{k}%
(\Omega)E_{i}(\Omega)$ and reads%

\begin{eqnarray}
H_{k}(\Omega)   =(-1)^{k}\exp\left(  -j2(k+2)\Omega\dfrac{d}{L}\right)
\nonumber\\
\times\dfrac{r-rg^{2}\exp(-2j\Omega)}{1-g^{2}r^{2}\exp(-2j\Omega
)}\left[ \dfrac{t^{2}g\exp\left(  -j\Omega\right)  }{1-g^{2}r^{2}%
\exp\left(  -2j\Omega\right)  }\right]  ^{k},\label{Hk}%
\end{eqnarray}
for $k=1\dots N-1$, and
\begin{align}
H_{N}(\Omega)  & =(-1)^{N}\exp\left(  -j2(N+1)\Omega\dfrac{d}{L}\right)
\nonumber\\
& \times\left[  \dfrac{t^{2}g\exp\left(  -j\Omega\right)  }{1-g^{2}r^{2}%
\exp\left(  -j2\Omega\right)  }\right]  ^{N}.\label{HN}%
\end{align}

{The relative simplicity of the transfer functions (\ref{Hk}) and
(\ref{HN}) arises from the zig-zag topology, which brings about the unique
properties of this structure.} If an additional ring is placed loading the
last output, Eq. \eqref{Hk} holds also for the $k=N$ case if multiplied by $r^{-1}$.  From Eqs. \eqref{Hk} and \eqref{HN}, we find that,
except for the first exponential term corresponding to the delay in coupling
sections, the response is periodic in $\Omega$ with $2\pi$ period. In the
following, we will restrict the analysis to the first period of the system
response. Also, the system is stable provided that the condition
\begin{equation}
gr<1
\end{equation}
is fulfilled. 

The total transmission group delay, calculated as $\tau_{g}=-\frac
{d\arg{H(\omega)}}{d\omega}$, is
\begin{align}
\dfrac{\tau_{g,k}}{\tau_{0}} & =2(k+2)\dfrac{d}{L}-\dfrac{2g^{2}(\cos
(2\Omega)-g^{2})}{1-2g^{2}\cos(2\Omega)+g^{4}}\nonumber\\
&   +k+(k+1)\dfrac{2g^{2}r^{2}(\cos(2\Omega)-g^{2}r^{2})}{1-2g^{2}r^{2}%
\cos(2\Omega)+g^{4}r^{4}}%
\end{align}
for $k=1\dots N-1$ and
\begin{equation}
\dfrac{\tau_{g,N}}{\tau_{0}}=2(N+1)\dfrac{d}{L}+N\left( 1+\dfrac{2g^{2}%
r^{2}(\cos(2\Omega)-g^{2}r^{2})}{1-2g^{2}r^{2}\cos(2\Omega)+g^{4}r^{4}}\right)
.
\end{equation}

If we tune the system to $\Omega=\pi/2$, the group delay is
\begin{equation}
\dfrac{\tau_{g,k}}{\tau_{0}}=2(k+2)\dfrac{d}{L}+\dfrac{2g^{2}}{1+g^{2}%
}+k-(k+1)\dfrac{2g^{2}r^{2}}{1+g^{2}r^{2}}%
\end{equation}
for $k=1\dots N-1$ and
\begin{equation}
\dfrac{\tau_{g,N}}{\tau_{0}}=2(N+1)\dfrac{d}{L}+N\left(  1-\dfrac{2g^{2}r^{2}%
}{1+g^{2}r^{2}}\right)  .
\end{equation}
If we take the limit $gr\rightarrow1$ and $d/L\rightarrow0$, the
total group delay vanishes $\tau_{g,N}=0$ and
\begin{equation}
\tau_{g,k}=\tau_{r}=\tau_{0}\dfrac{g^{2}-1}{g^{2}+1}.
\end{equation}
The input signal is transferred simultaneously to all system outputs 
with a common delay $\tau_{r}$. This behavior is what we call ubiquitous 
light transmission. The required double
limit implies neglecting the physical delay at the coupling sections 
and that the system is critically operated at the instability point. This 
limit can be approximated in a more
{realistic} way giving nearly simultaneous replicas of the input signal
below the instability threshold. The relative group delay accumulated between
consecutive {outputs} $\Delta\tau_{g}\equiv\tau_{g,k+1}-\tau_{g,k}$ in
this case is given by
\begin{equation}
\dfrac{\Delta\tau_{g}}{\tau_{0}}=\dfrac{2d}{L}+1-\dfrac{2g^{2}r^{2}}%
{1+g^{2}r^{2}},
\end{equation}
which is a monotonically decreasing function of the product $gr$ in the range
$0<gr<1$ and vanishes at $gr=1$ if the term $2d/L$ is neglected. For instance,
for $gr=0.9$ $\Delta\tau\simeq0.10\tau_{0}$ and for $gr=0.8$ $\Delta\tau
\simeq0.22\tau_{0}$.

\begin{figure}[tb]
\centerline{\includegraphics[width=8cm]{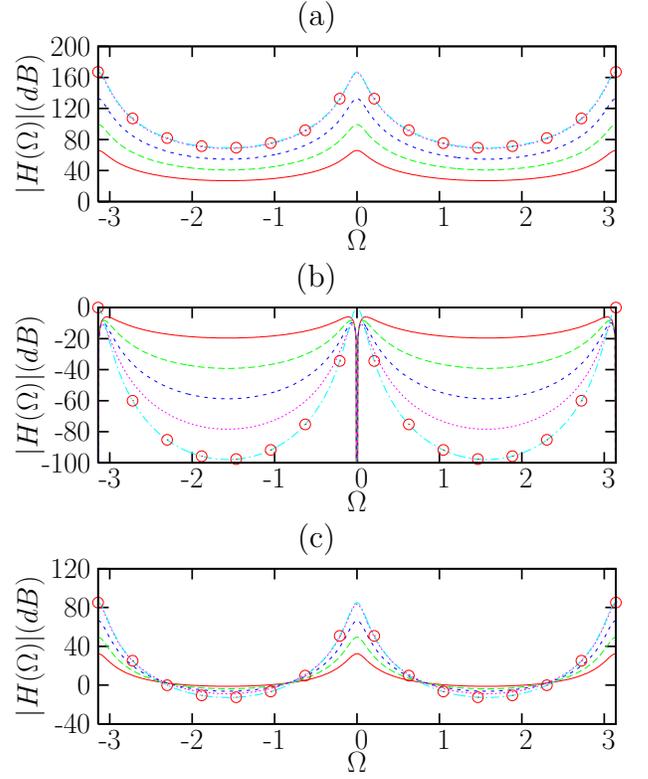}}\caption{ Amplitude
transfer function for $N=5$ and $rg=0.9$. (a) $r=0.1$, (b) $r=0.9$ and (c)
$r=0.5$. } \label{amplitud}
\end{figure}

\begin{figure}[tb]
\centerline{\includegraphics[width=8cm]{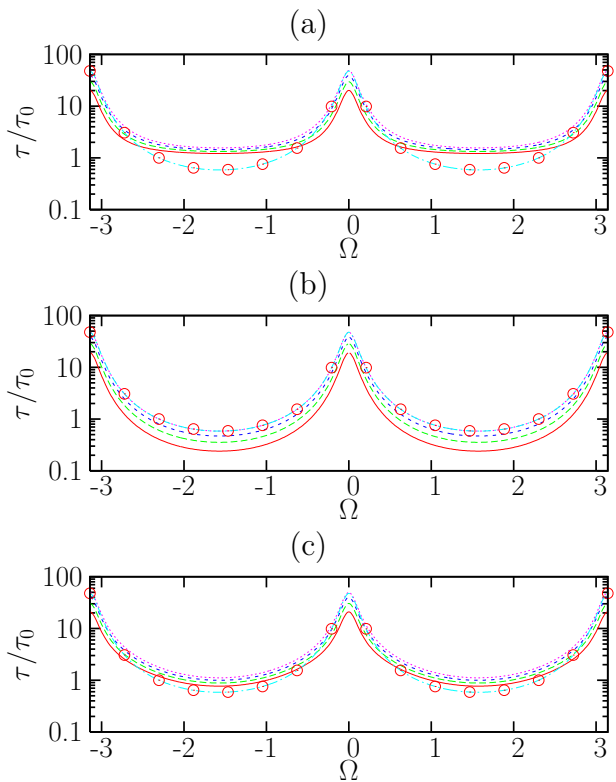}}\caption{ Transmission
group delay for $N=5$ and $rg=0.9$. (a) $r=0.1$, (b) $r=0.9$ and (c) $r=0.5$.} \label{retardo}
\end{figure}

Figures \ref{amplitud} and \ref{retardo} display the amplitude transfer function and the transmission
group delay, respectively, for a structure with $rg=0.9$. In these and
all the subsequent results presented, we assume a small value of $d/L=0.005$, $N=5$ and solid, long dashed, short dashed and dotted lines correspond to the 1st, 2nd, 3rd and 4th output, respectively, while the distinct 5th output is identified with small circles. The same value of $rg=0.9$ is 
considered in all the three cases 
of figures \ref{amplitud} and \ref{retardo} with $g=9$
(passive, lossy) in $(a)$, $g=1$ (passive, lossless) in $(b)$ and $g=1.8$
(active) in $(c)$. In the passive lossless case, $g=1$, the amplitude of the
$k$-th output vanishes, $k=1,\dots,N-1$, for the resonances at $\Omega=m\pi$
($m$ integer), while it is maximum for $N$-th output. For $g\neq1$, the output
amplitude of all the channels is maximum at the resonances. For the active
case $(a)$ a very flat group delay response over a broad bandwidth is found.
The amplitude responses in cases $(a)$ and $(b)$ show extreme variations in
the amplification or attenuation factors for the different outputs.

\begin{figure}[tbh]
\centerline{\includegraphics[width=8cm]{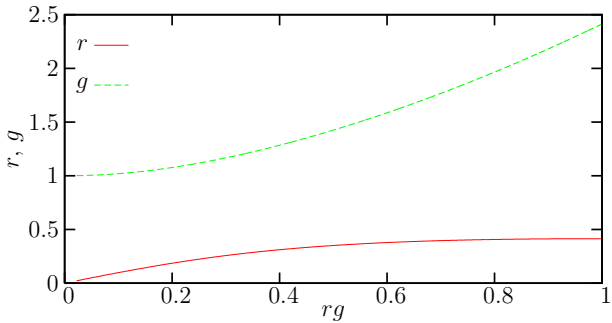}}\caption{ Values of $r$ and $g$ in the range $0<rg<1$ which produce equal amplitude outputs at $\Omega=\pm\pi/2$}\label{iguales}
\end{figure}

\begin{figure}[tbh]
\centerline{\includegraphics[width=8cm]{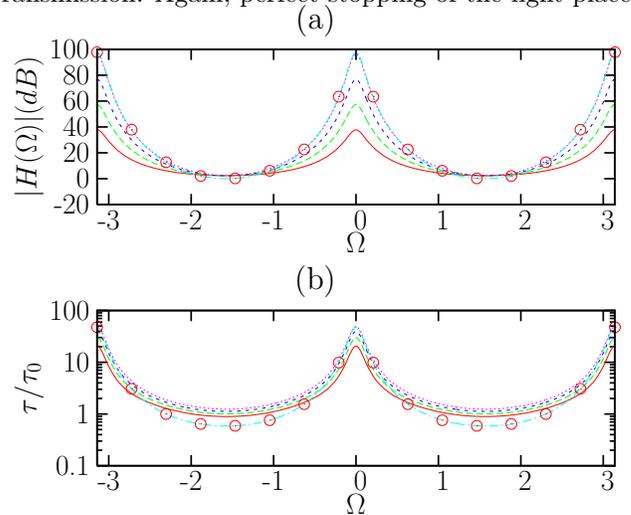}}\caption{ Amplitude
transfer function (a) and group delay (b) for $N=5$, $rg=0.9$ and $r$ and $g$ satisfying Equation \eqref{condicion-iguales}. }\label{respuesta-iguales}
\end{figure}

If the condition $g(1-r^{2})=1+r^{2}g^{2}$ is satisfied, the system response
for all the outputs $k=1\dots N-1$ at $\Omega=\pm{\pi}/{2}$ takes the
same value $\left| H_{k}\right| ={r(1+g^{2})}{\left( 1+r^{2}g^{2}\right)^{-1}}$ and $\left| H_{N}\right| =1$. Figure \ref{iguales} shows the values of $r$ and $g$ satisfying this relation in the stability range $0<rg<1$.  It is required that the waveguides show amplification, $g>1$, and
\begin{equation}
r=\sqrt{\dfrac{g-1}{g(g+1)}}\label{condicion-iguales}.
\end{equation}
Figure \ref{respuesta-iguales} displays the amplitude transfer functions and group delays for the various outputs of a $N=5$ structure and $rg=0.9$ when the condition given in Equation \eqref{condicion-iguales} is satisfied.

We now turn to the slow light transmission in the system.  If we keep the system parameters fixed, but tune the input to one of the system
resonances, $\Omega=0$. The net group delay is
\begin{equation}
\dfrac{\tau_{g,k}}{\tau_{0}}=2(k+2)\dfrac{d}{L}+h+k+(k+1)\dfrac{2g^{2}r^{2}}{1-g^{2}r^{2}},%
\end{equation}
$k=1\dots N-1$, where $h=1$ for $g=1$ and $h=2g^2(g^2-1)^{-1}$ otherwise.  For the N-th output, 
\begin{equation}
\dfrac{\tau_{g,N}}{\tau_{0}}=2(N+1)\dfrac{d}{L}+N+N\dfrac{2g^{2}r^{2}}%
{1-g^{2}r^{2}},
\end{equation}
with $\tau_{g,k}\rightarrow\infty$ for $k=1\dots N$ when $gr\rightarrow1$,
providing a regime for the slowing down of the optical signal transmission.
Again, perfect {stopping} of the light places the operation point
critically at the border of the instability region and smaller values of the
product $gr$ should be considered as more {practical} conditions. The
{relative} delay between two adjacent outputs in this case is
\begin{equation}
\dfrac{\Delta\tau_{g}}{\tau_{0}}=\dfrac{2d}{L}+1+\dfrac{2g^{2}r^{2}}%
{1-g^{2}r^{2}}.
\end{equation}
If we neglect the $2d/L$ term, this relative delay is 
$\Delta\tau\simeq 9.5\tau_0 $ for $gr=0.9$ and 
$\Delta\tau\simeq 19.5\tau_0$ for $gr=0.95$.

Light propagation in a ZZRR chain has been analyzed. The system exhibits both
slow and fast {light} transmission. In the fast case, a novel type of
superluminal propagation, ubiquitous transmission, has been identified and
studied. The novel {effects} addressed in this Letter could find also
new applications in optical integrated circuits where ubiquitous signal
transmission can be used for in-chip
{synchronization}. Using previously described {tuning}
{mechanisms} \cite{Blair2006,Liu2008} one can obtain a relative delay
between outputs which can be controlled in a very broad margin. This can be
used, for instance, to {obtain} a tunable system for pulse repetition
rate multiplication \cite{Preciado2008} by combining various outputs.

This work has been funded by the Spanish MCINN and FEDER, project number
TEC2007-67429-C02-01and 02 and JCyL VA001A08.

\bibliographystyle{ol}
\bibliography{zigzag}

\begin{thebibliography}{1}
\newcommand{\enquote}[1]{``#1''}

\bibitem{Boyd2006}
R.~Boyd, D.J.Gauthier, and A.~Gaeta, Opt.\ Photon.\ News \textbf{17}, 18
  (2006).

\bibitem{Parra2007}
E.~Parra and J.~Lowell, Opt.\ Photon.\ News \textbf{18}, 40 (2007).

\bibitem{Heebner2002}
J.~Heebner and R.~Boyd, J.\ Mod.\ Opt.\ \textbf{49}, 2629 (2002).

\bibitem{Fraile2007}
F.~Fraile-Pelaez and P.~Chamorro-Posada, Opt.\ Express \textbf{15}, 3177
  (2007).

\bibitem{Liu2008}
F.~Liu, Q.~Li, Z.~Zhang, M.~Qiu, and Y.~Su, IEEE J. Select.\ Top.\ Quantum
  Electron.\ \textbf{14}, 706 (2008).

\bibitem{Xia2007}
F.~Xia, L.~Sekaric, and Y.~Vlasov, Nat.\ Photon.\ \textbf{1}, 65 (2007).

\bibitem{Scheuer2005}
J.~Scheuer, G.~Paloczi, J.~Poon, and A.~Yariv, Opt.\ Photon.\ News \textbf{16},
  36 (2005).

\bibitem{Blair2006}
S.~Blair and K.~Zheng, Opt.\ Express \textbf{16}, 11162 (2006).

\bibitem{Preciado2008}
M.~Preciado and M.~Muriel, Opt.\ Express \textbf{16}, 11162 (2008).

\end{thebibliography}
\end{document}